\documentclass[sigconf,nonacm]{acmart}
\usepackage{booktabs}\usepackage{colortbl}\usepackage{tabularx}\newcolumntype{C}{>{\centering\arraybackslash}X}\newcolumntype{R}{>{\raggedleft\arraybackslash}X}\usepackage{multirow}\usepackage{stmaryrd}\usepackage{xcolor}
\newcommand{\cnum}[1]{\textcircled{\scriptsize #1}}

\setcopyright{none}\renewcommand\footnotetextcopyrightpermission[1]{}
\begin{document}
\title{VeilTune: Secure Fine-Tuning of Private LoRA Adapters on Public Pretrained Language Models}
\author{Huiyu Xie}
\orcid{0009-0001-4752-1247}
\affiliation{\institution{Information Engineering University}\city{Zhengzhou}\postcode{450002}\country{China}}
\affiliation{\institution{Key Laboratory of Cyberspace Endogenous Safety \& Security of Henan Province}\city{Zhengzhou}\postcode{450000}\country{China}}
\email{xiehuiyuaaa@163.com}
\author{Tanping Zhou}
\orcid{0000-0001-6325-5965}
\affiliation{\institution{Chinese People's Armed Police Force Engineering University}\city{Xi'an}\country{China}}
\email{tanping2020@iscas.ac.cn}
\author{Tao Hu}
\orcid{0000-0001-7641-5622}
\affiliation{\institution{Information Engineering University}\city{Zhengzhou}\postcode{450002}\country{China}}
\affiliation{\institution{Key Laboratory of Cyberspace Endogenous Safety \& Security of Henan Province}\city{Zhengzhou}\postcode{450000}\country{China}}
\email{hutaondsc@163.com}
\author{Xinlei Liu}
\orcid{0000-0001-6586-4457}
\affiliation{\institution{Information Engineering University}\city{Zhengzhou}\postcode{450002}\country{China}}
\affiliation{\institution{Key Laboratory of Cyberspace Endogenous Safety \& Security of Henan Province}\city{Zhengzhou}\postcode{450000}\country{China}}
\email{liuxl@stu.ouc.edu.cn}
\author{Hailong Ma}
\authornote{Corresponding author.}
\orcid{0000-0003-2497-366X}
\affiliation{\institution{Information Engineering University}\city{Zhengzhou}\postcode{450002}\country{China}}
\affiliation{\institution{Key Laboratory of Cyberspace Endogenous Safety \& Security of Henan Province}\city{Zhengzhou}\postcode{450000}\country{China}}
\email{longmanclear@163.com}
\renewcommand{\shortauthors}{Xie et al.}

\begin{abstract}
Fine-tuning public pretrained language models with lightweight adapters such as LoRA has become the mainstream way to adapt them to downstream tasks. A service provider often trains its adapter on a data owner's samples, so two assets need protection: the data owner's samples and the provider's adapter. Existing secure fine-tuning protocols protect the samples but reveal to the data owner intermediate results of every training step, such as the hidden representations or logits of its inputs. We show that a single step suffices to copy the adapter: from the hidden representations of one mini-batch of 32 samples, the data owner fits an adapter that agrees with the provider's on 99.2\% of test inputs, and logits alone yield about 94\%. Protecting both assets therefore requires that no value that depends on the adapter be revealed throughout training. We present VeilTune, to the best of our knowledge the first protocol that fine-tunes a pretrained language model with backpropagation entirely in secure multi-party computation (MPC). VeilTune computes only what the loss depends on, replaces generic MPC operators with training-aware nonlinear operators that stay accurate on the values that training produces, and obliviously skips the steps corrupted by truncation failures. On four few-shot tasks with RoBERTa-large, VeilTune nearly matches plaintext training and is 5.3 points more accurate on average than SecP-Tuning, the state-of-the-art secure fine-tuning protocol. On SST-2, it achieves a $3.5\times$ end-to-end speedup and a $3.6\times$ reduction in communication over SecP-Tuning, and $26\times$ and $17\times$ over full fine-tuning in MPC.

\end{abstract}
\maketitle
\section{Introduction}
\label{sec:intro}

Large language models have become a building block of many Web services, and most services adapt a public pretrained model to their task by training a small adapter, such as LoRA~\cite{lora}, rather than the whole model.
The adapter is then often trained by a service provider on data that belongs to someone else.
Such training involves two assets that the parties cannot share: the provider's adapter, which encodes its know-how and is a business asset, and the data owner's labeled samples, which are private.
What the two parties need is therefore \emph{two-way privacy}: the provider must not see the data, and the data owner must not see the adapter at any time.

Existing secure fine-tuning protocols~\cite{sap,privtuner,secptuning} protect the data, but they open intermediate results of every training step to the data owner, such as the hidden representations or logits of its inputs, on the grounds that none of them is the adapter itself.
We show that these results nevertheless reveal the adapter (Section~\ref{sec:attack}).
Every opened value that depends on the adapter answers a query to it, so the data owner can collect the answers and fit its own copy by model extraction.
From the hidden representations of a single mini-batch of 32 samples, our adapter extraction attack obtains a copy that agrees with the provider's adapter on 99.2\% of test inputs, and from logits alone on about 94\%, even when the data owner follows the protocol.
Two-way privacy therefore requires that no value that depends on the adapter be revealed during training, including the gradients and the optimizer state, so that backpropagation and the parameter update must run in protected form as well.

This leads to the question that this paper answers: \emph{can a pretrained language model be fine-tuned with backpropagation entirely in secure multi-party computation (MPC), at a practical cost and with plaintext accuracy?}
No existing protocol does so: SAP~\cite{sap} and PrivTuner~\cite{privtuner} backpropagate from values opened to the data owner, and SecP-Tuning~\cite{secptuning} avoids backpropagation altogether.
Avoiding it is no way out, since forward-only, zeroth-order training needs about 16 times more steps and 24 times more communication to converge (Section~\ref{sec:eval:zo}).
Running backpropagation in MPC, however, faces two obstacles.
The first is cost: in MPC every nonlinear operator becomes an interactive protocol, and training adds the derivative of every operator and the optimizer on top of the forward pass.
The second is accuracy: training produces values far larger and far smaller than generic MPC operators and the fixed-point encoding can handle, and an error that would spoil a single prediction in inference is stored in the optimizer state and distorts every later update.

To overcome these obstacles, we present VeilTune, to the best of our knowledge the first protocol that fine-tunes a pretrained language model with backpropagation entirely in MPC.
VeilTune runs the whole training step in MPC and reveals only the trained adapter, to the provider alone, and makes this practical and accurate with three techniques.
\emph{Token packing} computes only what the loss depends on: since the token lengths and \texttt{[MASK]} positions of the examples are public, VeilTune drops padding and evaluates the last block only at the \texttt{[MASK]} position, without revealing anything new.
\emph{Training-aware nonlinear operators} are accurate for inputs of any magnitude, because they approximate each function only on a small interval and reduce every other input to an exact formula or to that interval; they also need less communication than generic MPC operators.
\emph{Oblivious step skipping} recognizes a step corrupted by a truncation failure from the magnitude of its gradient and discards it through a secret bit, so that neither party learns which steps were skipped, while the accuracy stays close to that of plaintext training.

\noindent In summary, we make the following contributions:
\begin{itemize}
\item \emph{Adapter extraction attack.} We show that the intermediate results opened by SAP, PrivTuner and SecP-Tuning let the data owner copy the adapter: a single training step yields an agreement of 99.2\% from hidden representations and about 94\% from logits.
\item \emph{Secure backpropagation.} We design VeilTune, which provides two-way privacy by running the whole training step in MPC, and introduce token packing, training-aware nonlinear operators and oblivious step skipping to make it practical and accurate.
\item \emph{Evaluation.} With RoBERTa-large on four few-shot tasks, VeilTune nearly matches plaintext training, staying within 0.3 to 1.7 points of it, and is 5.3 points more accurate than SecP-Tuning on average. It is also more efficient: on SST-2, VeilTune achieves a $3.5\times$ end-to-end speedup and a $3.6\times$ reduction in communication over SecP-Tuning, and $26\times$ and $17\times$ over full fine-tuning in MPC.
\end{itemize}

\subsection{Related Work}
\label{sec:related}

\paragraph{Secure fine-tuning.}
Existing protocols for fine-tuning language models on private data keep the data from the provider, but leave the adapter exposed to the data owner.
In some, the adapter belongs to the data owner by design: in Offsite-Tuning~\cite{offsite}, the data owner trains it on a compressed copy of the model, and in the fully homomorphic fine-tuning of Rho et al.~\cite{rho}, it is encrypted under the data owner's key.
Others keep the adapter with the provider and let it train without seeing the data: SAP~\cite{sap} leaves the first layers of the model to the data owner, PrivTuner~\cite{privtuner} evaluates the model on homomorphically encrypted data, and SecP-Tuning~\cite{secptuning} evaluates it in MPC.
Yet each of them hands the data owner hidden representations or predictions computed from the adapter, which suffice to copy it (Section~\ref{sec:attack}).

\paragraph{Secure Transformer inference.}
A large body of work evaluates Transformers on private inputs, with two-party protocols that combine homomorphic encryption and secret sharing~\cite{iron,bolt,bumblebee,nimbus,ciphergpt}, function secret sharing~\cite{sigma}, or three parties~\cite{puma}, and with models retrained to be friendlier to secure computation~\cite{mpcformer}.
Their approximations of GELU, softmax and LayerNorm are designed for the forward pass on the activations of inference.
Training additionally needs the derivative of every operator, must handle values that only training produces, such as Adam's second moments, and keeps an optimizer state over hundreds of steps, in which a rare numerical failure is not averaged away but accumulates (Section~\ref{sec:design:challenges}).
These operators therefore cannot be reused for training as they are, which is why VeilTune designs its nonlinear operators for training (Section~\ref{sec:design:kernels}).

\paragraph{Secure training frameworks.}
General frameworks for secure training, such as SecureML~\cite{secureml}, Falcon~\cite{falcon}, CrypTen~\cite{crypten} and Piranha~\cite{piranha}, train networks from scratch, mostly image classifiers.
Fine-tuning a pretrained language model is harder for fixed-point arithmetic, because the outlier dimensions of the model and the small second moments of Adam fall outside the range that it represents accurately (Section~\ref{sec:design:challenges}), so these frameworks alone do not support such fine-tuning.

\section{Preliminaries}
\label{sec:prelim}

\subsection{Fine-Tuning With LoRA}
\label{sec:prelim:ft}

\paragraph{Prompt-based classification.}
Following LM-BFF~\cite{lmbff}, we cast classification as masked language modeling.
The input is wrapped in a template that contains one \texttt{[MASK]} token, such as ``\textit{$\langle$sentence$\rangle$} It was \texttt{[MASK]}.'', and each class $c$ is assigned a label word.
The logit $z_c$ is the score that the masked-language-model head gives to the label word of $c$ at the \texttt{[MASK]} position, and training minimizes the cross-entropy $\ell=-\log p_y$ for the label $y$, where $p=\mathrm{softmax}(z)$.
Since the head reads only the \texttt{[MASK]} position, the other positions affect the loss only through attention.

\paragraph{LoRA.}
LoRA~\cite{lora} freezes the pretrained weights $W$ and adds to a projection $hW$ a low-rank branch,
\begin{equation}
hW \;\mapsto\; hW + hA B,\qquad A\in\mathbb{R}^{d\times r},\ B\in\mathbb{R}^{r\times d},\ r\ll d,
\label{eq:lora}
\end{equation}
where $d$ is the hidden dimension and $B$ is initialized to zero, so that the branch starts as a no-op.
Only the adapter $\theta=\{A_j,B_j\}$ of the adapted layers $j$ is trained.

\paragraph{Adam.}
Adam~\cite{kingma2015adam} keeps a first moment $m$ and a second moment $v$ for every parameter and, given the gradient $g$ at step $t$, updates
\begin{equation}
\begin{aligned}
&m\leftarrow\beta_1 m+(1-\beta_1)\,g,\qquad v\leftarrow\beta_2 v+(1-\beta_2)\,g^2,\\
&\theta\leftarrow\theta-\eta\,\frac{m/(1-\beta_1^t)}{\sqrt{v/(1-\beta_2^t)}+\epsilon},
\end{aligned}
\label{eq:adam-std}
\end{equation}
where $\beta_1,\beta_2$ are decay rates, $\eta$ is the learning rate, and the divisions by $1-\beta_1^t$ and $1-\beta_2^t$ correct the bias of the zero-initialized moments.

\subsection{Secure Two-Party Computation}
\label{sec:prelim:mpc}

\paragraph{Sharing and arithmetic.}
We use the standard two-party arithmetic over a ring~\cite{crypten}.
A real value $x$ is encoded in fixed point as $\lfloor x\cdot 2^{f}\rceil$ with $f$ fractional bits and additively shared over $\mathbb{Z}_{2^{64}}$, so that $\llbracket x\rrbracket_0+\llbracket x\rrbracket_1 = x\cdot 2^{f} \bmod 2^{64}$.
Additions and products with public values need no communication; we call such operations local.
A product of two shared values consumes a Beaver triple~\cite{beaver}, generated in the preprocessing phase (Section~\ref{sec:setting:threat}), and one round of communication.
A comparison $[x<0]$ converts the sharing into shares of the individual bits and evaluates a Boolean circuit on them, in a number of rounds logarithmic in the bit width.
Operations on many entries are batched into one call, so the number of rounds depends on the depth of a computation rather than on its width.

\paragraph{Truncation and its failures.}
After a product the encoding carries $2f$ fractional bits, and each party truncates its share back to $f$ bits locally~\cite{secureml}.
This is free but fails with a small probability: if the two shares wrap around the ring, the result is off by about $2^{32}$.
For a product whose true value is $x$, the failure probability is approximately
\begin{equation}
\Pr[\text{failure}] \approx \frac{|x|\cdot 2^{2f}}{2^{63}} = |x|\cdot 2^{-31},
\label{eq:truncfail}
\end{equation}
so larger products fail more often.
This estimate follows the analysis of local truncation in SecureML~\cite{secureml} and holds whenever $|x|\cdot 2^{2f}\ll 2^{63}$.
Exact truncation protocols~\cite{sigma} avoid such failures, but add communication to every product; VeilTune keeps local truncation and instead tolerates its failures (Section~\ref{sec:design:adam}).

\paragraph{Comparisons on fewer bits.}
The communication of a comparison grows with the bit width of the shares.
When $x$ is known to be small, the comparison can therefore run on shorter shares: each party keeps only the lowest $k$ bits of its share, which needs no communication, and the two shortened shares add up to the encoding of $x$ modulo $2^{k}$.
If the encoding fits in $k$ bits including its sign, that is, $|x|\cdot2^{f}<2^{k-1}$, this sum has the same sign as $x$, so $[x<0]$ can be computed on $k$-bit instead of 64-bit shares.
VeilTune uses this for comparisons whose inputs are bounded (Section~\ref{sec:design:kernels}): on 32-bit and 16-bit shares, a comparison needs less than half and about a fifth of the communication on 64-bit shares (Appendix~\ref{app:kernels}).

\section{Problem Setting}
\label{sec:setting}

\subsection{System and Threat Model}
\label{sec:setting:threat}

\paragraph{Parties.}
The \emph{data owner} $P_0$ holds a labeled dataset $\mathcal{D}$ of examples $(x_i,y_i)$, whose inputs and labels are private.
The \emph{model owner} $P_1$ holds a task adapter, the LoRA parameters $\theta$ on the query and value projections of the upper Transformer blocks, which may be freshly initialized or already trained on $P_1$'s own data; in either case $P_1$ wants it hidden from $P_0$ at all times.
The base weights $W$ of an open-weight model such as RoBERTa~\cite{roberta}, the protocol, the training configuration and the shape and placement of the adapter are public; the data owner also has public unlabeled text from the task domain.
The parties fine-tune $\theta$ on $\mathcal{D}$ for $T$ steps and give the result $\theta_T$ to $P_1$.

\paragraph{Adversary.}
We consider a static, computationally bounded adversary that corrupts one of the two parties and is semi-honest~\cite{pencil,secptuning}: the corrupted party follows the protocol but tries to infer the other party's secrets from everything it observes.
Following the protocol still leaves a party free to choose its own inputs, so the data owner may, for instance, add public sentences to its mini-batches as additional queries.
As in prior work~\cite{crypten,secptuning}, input-independent correlated randomness such as Beaver triples~\cite{beaver} is supplied in a preprocessing phase by a dealer that colludes with neither party.

\paragraph{Goals.}
A corrupted data owner tries to obtain a copy of the adapter, that is, a model whose predictions agree with those of $\theta_T$ on test inputs (Section~\ref{sec:attack}); a corrupted model owner tries to infer the training inputs or labels.
Attacks that deviate from the protocol, side channels, and attacks that need only the released $\theta_T$ are out of scope.

\subsection{Security Goal}
\label{sec:setting:goal}

Our goal is \emph{two-way privacy} against the adversary above: a corrupted data owner learns nothing about the adapter, and a corrupted model owner learns nothing about the training data beyond the trained adapter $\theta_T$ that it receives.
Following the simulation paradigm~\cite{goldreich,canetti}, we capture it by an ideal functionality $\mathcal{F}_{\textsf{FT}}$ that receives $\mathcal{D}$ from $P_0$ and $\theta_0$ from $P_1$, runs $T$ training steps internally, and returns $\theta_T$ to $P_1$ and nothing to $P_0$, and by a leakage profile $\mathcal{L}(\mathcal{D})$ that consists of the public model, the training configuration, and the token length and \texttt{[MASK]} position of each example.

\begin{definition}[Two-way privacy]
\label{def:security}
A protocol $\Pi$ securely realizes $\mathcal{F}_{\textsf{FT}}$ with leakage $\mathcal{L}$ against a static semi-honest adversary if there exist probabilistic polynomial-time simulators $\mathcal{S}_0$ and $\mathcal{S}_1$ such that for all $\mathcal{D}$ and $\theta_0$,
\begin{align*}
\mathcal{S}_0\bigl(\mathcal{D},\mathcal{L}(\mathcal{D})\bigr) &\approx_c \mathsf{view}^{\Pi}_{0}(\mathcal{D},\theta_0),\\
\mathcal{S}_1\bigl(\theta_0,\theta_T,\mathcal{L}(\mathcal{D})\bigr) &\approx_c \mathsf{view}^{\Pi}_{1}(\mathcal{D},\theta_0),
\end{align*}
where $\mathsf{view}^{\Pi}_{i}$ is the view of $P_i$ in an execution of $\Pi$ and $\theta_T$ is the output of $\mathcal{F}_{\textsf{FT}}$.
\end{definition}

In particular, no value that depends on the adapter reaches the data owner during training, so the extraction attack of Section~\ref{sec:attack} has nothing to query.

\section{Adapter Extraction Attacks on Existing Protocols}
\label{sec:attack}

In this section, we show that the intermediate results opened by existing secure fine-tuning protocols are enough for the data owner to copy the model owner's adapter.
We mount this attack on SAP~\cite{sap}, PrivTuner~\cite{privtuner} and SecP-Tuning~\cite{secptuning}, which all keep the model owner's model hidden from the data owner and represent the three main designs: split learning, homomorphic encryption with backpropagation, and MPC with derivative-free optimization.
SAP opens hidden representations, whereas PrivTuner and SecP-Tuning open only the logits on the data owner's inputs.

\subsection{Adapter Extraction Attack}
\label{sec:attack:setting}

Any opened value that depends on the current adapter $\theta_t$ and is computed on inputs supplied by the data owner answers a query to the adapter.
Opening less makes each answer less informative, but does not remove the answers.
We therefore consider an \emph{adapter extraction attack} (AEA), in which the data owner collects these answers and fits its own copy of the adapter by model extraction~\cite{tramer2016stealing,orekondy2019knockoff}.

\paragraph{Attacker knowledge.}
Besides the knowledge of Section~\ref{sec:setting:threat}, the data owner sees the values opened at each step, $o_t=g(X_t;\theta_t)$ on its mini-batch $X_t$, where $g$ returns the hidden representations at the output of the adapted blocks for SAP and the logits for PrivTuner and SecP-Tuning.

\paragraph{Attack method.}
The attacker initializes an adapter $\hat\theta$ of the same structure and minimizes $\sum_t\lVert g(X_t;\hat\theta)-o_t\rVert^2$ over the last $K$ observed steps, matching hidden representations under SAP and logits under PrivTuner and SecP-Tuning.
We evaluate a \emph{passive} attacker, which uses only its training mini-batches, and an \emph{active} one, which also places public sentences in its mini-batches as additional queries; these sentences do not enter the training loss, so the attacker remains semi-honest (Section~\ref{sec:setting:threat}).

\paragraph{Metric.}
We measure leakage by the \emph{agreement} of $\hat\theta$ with the trained adapter $\theta_T$, the fraction of test inputs on which both predict the same label~\cite{jagielski2020high}, with higher values indicating more leakage.
As a baseline, we report the agreement of a \emph{local model} that the data owner trains on its own examples without the protocol.

\begin{figure}[t]
\centering
\includegraphics[width=\columnwidth]{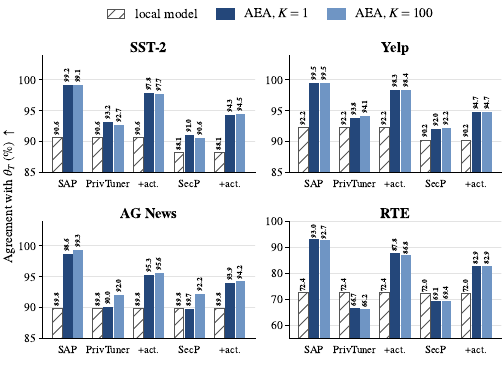}
\caption{Attack performance (agreement with $\theta_T$, \% $\uparrow$) of AEA against existing protocols, compared with the local model, on four tasks (mean of three seeds). ``+act.'' marks the active attacker, the other groups are passive; SecP is SecP-Tuning.}
\label{fig:leak}
\end{figure}

\subsection{Attack Results}
\label{sec:attack:results}

We attack the adapter of our main experiments, on RoBERTa-large~\cite{roberta}, on the four tasks of Section~\ref{sec:eval}; Appendix~\ref{app:leak} gives the setup and the full results.

\paragraph{Every protocol leaks the adapter.}
Under every protocol and task, the stronger of the two attackers extracts an adapter that agrees with $\theta_T$ more often than the local model does (Figure~\ref{fig:leak}): on SST-2, Yelp and AG News on at least 93.9\% of the test inputs, against at most 92.2\% for the local model, and on RTE, the hardest task, on 82.9\% to 93.0\%, against at most 72.4\%.
The extracted adapter is also close to $\theta_T$ in accuracy, so the data owner obtains a usable copy.
A single step of opened values already suffices: extracting from the last 100 steps instead raises the agreement by at most 2.5 points, so opening values rarely or only late in training is no remedy.

\paragraph{The more the opened values reveal, the more the adapter leaks.}
Hidden representations give the adapter away at once: under SAP, even the passive attacker reaches 99.2\% agreement on SST-2 and 99.5\% on Yelp.
Logits reveal less, so the passive attacker gains little or nothing under PrivTuner and SecP-Tuning, but the active attacker turns them into enough queries to recover the adapter.
SecP-Tuning leaks least, because its logits come from perturbed candidates rather than from $\theta_t$.

Our attack does not break the security proofs of these protocols, which allow the opened values to be revealed; it shows that revealing them is already enough to copy the adapter.
A protocol that protects the adapter must therefore keep every value that depends on it hidden throughout training.

\section{VeilTune Design}
\label{sec:design}

In this section, we introduce the design of VeilTune.
Section~\ref{sec:attack} showed that a protocol that protects the adapter must reveal to the data owner no value that depends on it.
VeilTune meets this requirement with MPC: $P_0$ and $P_1$ run the whole training step, including the forward pass, backpropagation and the Adam update, on secret shares, and open only $\theta_T$ to $P_1$.
We first give an overview of the protocol (Section~\ref{sec:design:overview}) and identify the challenges of training in MPC (Section~\ref{sec:design:challenges}).
We then present token packing (Section~\ref{sec:design:layout}), training-aware nonlinear operators (Section~\ref{sec:design:kernels}) and oblivious step skipping (Section~\ref{sec:design:adam}), and finally analyze security (Section~\ref{sec:design:security}).

\subsection{Overview}
\label{sec:design:overview}

\begin{figure*}[t]
\centering
\includegraphics[width=\textwidth]{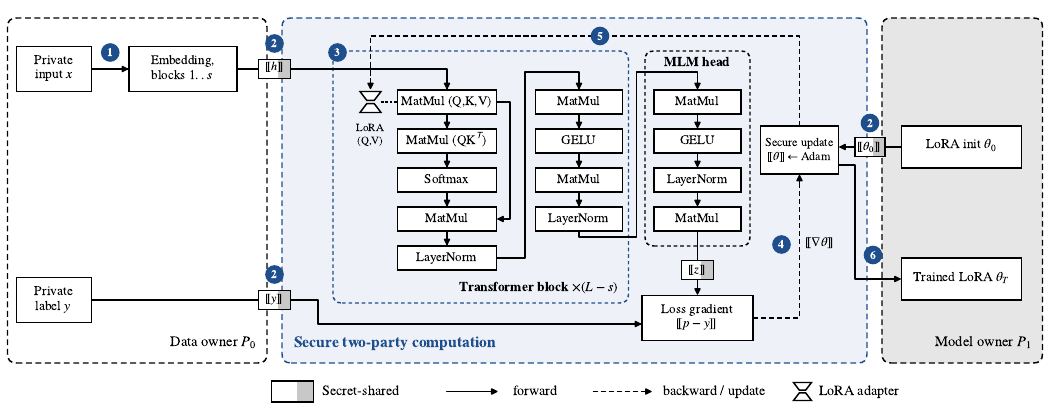}
\caption{Workflow of VeilTune and one secure training step. All values inside the middle box are secret-shared, and the Transformer block is repeated for the $L-s$ adapted blocks.}
\label{fig:workflow}
\end{figure*}

\paragraph{Workflow.}
Figure~\ref{fig:workflow} shows the workflow of VeilTune.
Let $L$ be the number of Transformer blocks and $s$ the split layer.
\cnum{1}~Blocks $1,\dots,s$ carry no adapter, so $P_0$ evaluates them locally and obtains the hidden states $h=\phi_{1:s}(x;W)$ in plaintext.
\cnum{2}~$P_0$ secret-shares $h$ and $y$, and $P_1$ secret-shares $\theta_0$.
The remaining blocks $s{+}1,\dots,L$, the masked-language-model head and the whole training procedure are then evaluated jointly on these shares, one mini-batch of $b$ examples per step.
\cnum{3}~The forward pass produces the logits $\llbracket z\rrbracket$.
\cnum{4}~Backpropagation computes the gradient of the classification loss with respect to $\theta$.
\cnum{5}~An Adam update produces the new adapter shares.
\cnum{6}~After $T$ steps the protocol opens the final adapter $\theta_T$ to $P_1$ only.
Table~\ref{tab:notation} in Appendix~\ref{app:notation} summarizes the notation.

\subsection{Challenges and Design Principles}
\label{sec:design:challenges}

\begin{figure}[t]
\centering
\includegraphics[width=\columnwidth]{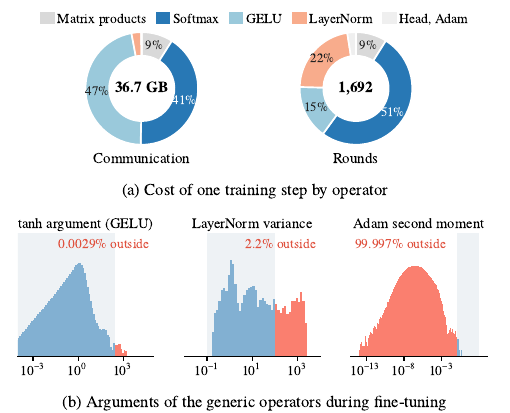}
\caption{A secure training step built from generic operators. (a)~Communication and rounds by operator. (b)~Arguments of three generic operators during fine-tuning; the shaded band is the range on which the operator is accurate.}
\label{fig:profile}
\end{figure}

Running the training step in MPC is both costly and error-prone.
To see why, Figure~\ref{fig:profile} profiles a training step built from \emph{generic operators}, the approximations of nonlinear functions that general-purpose MPC frameworks provide~\cite{crypten}.

\paragraph{Challenge 1: Nonlinear operators dominate the cost.}
Products with the public weights $W$ are local, so matrix products account for only 9\% of the communication; softmax, GELU and LayerNorm account for nearly all the rest (Figure~\ref{fig:profile}a).
Secure inference faces the same bottleneck~\cite{iron,bolt}, but training makes it worse: every nonlinear operator must also supply its derivative, and Adam adds an inverse square root for every adapter entry.

\paragraph{Challenge 2: Generic operators fail on the values that training produces.}
Each generic operator is accurate only on a bounded range of inputs, which covers typical activations but not the values that fine-tuning produces (Figure~\ref{fig:profile}b).
The inverse square root, for instance, is accurate only between about $0.1$ and $100$, yet LayerNorm variances exceed 100 because of the \emph{outlier dimensions} of pretrained Transformers~\cite{kovaleva2021bertbusters,dettmers2022llmint8}, and almost all second moments of Adam lie below $0.1$.
In inference an inaccurate operator spoils one prediction; in training its error enters the optimizer state and distorts every later update.

\paragraph{Design principles.}
Each challenge leads to one principle.

\emph{(i) Pay only for what training needs.}
No communication is spent on values that cannot affect the loss (Section~\ref{sec:design:layout}), and the nonlinear operators that remain keep their communication low enough for fine-tuning in MPC to be practical (Section~\ref{sec:design:kernels}).

\emph{(ii) Stay accurate on the values that training produces.}
VeilTune's operators are accurate on every input that training produces (Section~\ref{sec:design:kernels}), and an error that does occur must not persist across steps (Section~\ref{sec:design:adam}).

\subsection{Token Packing}
\label{sec:design:layout}

In plaintext, padding and unused outputs cost little; in MPC, every position that is computed costs communication, whether or not it affects the result.
Few-shot fine-tuning wastes many positions: the inputs are short and vary in length, and the classifier reads a single position of each.
Since the token lengths and \texttt{[MASK]} positions are public (Section~\ref{sec:setting:goal}), VeilTune can lay out the computation around them without revealing anything new, which it does in two ways.

\begin{figure}[t]
\centering
\includegraphics[width=0.88\columnwidth]{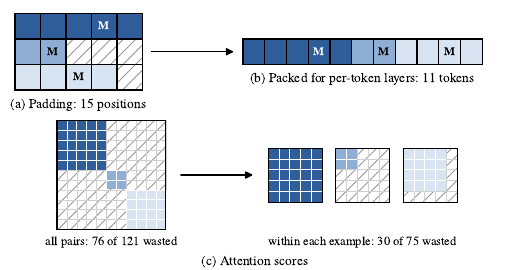}
\caption{Token packing for a batch of three examples. M marks the \texttt{[MASK]} token; hatched cells are computed but wasted. (a)~Padding. (b)~Packed tokens for the per-token layers. (c)~Attention scores over all pairs of packed tokens (left) and within each example after regrouping (right).}
\label{fig:pack}
\end{figure}

\paragraph{Packing tokens instead of padding.}
Padding a batch to its longest example makes every layer process $b\cdot\max_i n_i$ positions, of which only $N=\sum_i n_i$ are real tokens, and few-shot inputs are short and vary widely in length.
Token packing removes this waste: VeilTune concatenates the real tokens of a batch into one sequence of $N$ tokens, which the per-token layers, namely the projections, GELU and LayerNorm, process directly (Figure~\ref{fig:pack}b).
Attention is the only layer that mixes tokens, and scoring all pairs of packed tokens would spend most of the scores on pairs from different examples, which are discarded (Figure~\ref{fig:pack}c).
VeilTune therefore regroups the packed tokens by example and scores them within each example; the regrouping is local because the lengths are public.
Token packing nearly halves the communication of a step (Section~\ref{sec:eval:ablation}).
If the lengths must stay hidden, the inputs can instead be padded to a fixed public length.

\paragraph{Computing only the \texttt{[MASK]} row in the last block.}
In the last adapted block, only the output at the \texttt{[MASK]} position reaches the loss; the other tokens matter only as the keys and values that this position attends to.
VeilTune therefore computes the keys and values of every token in that block, but the query, the rest of the block and the classification head for the $b$ \texttt{[MASK]} rows only.
In our experiments, this removes a further 11\% of the communication of a step (Section~\ref{sec:eval:ablation}).

\subsection{Training-Aware Nonlinear Operators}
\label{sec:design:kernels}\label{sec:design:backward}

Existing secure operators, both the generic ones and those designed for secure inference~\cite{iron,bolt}, target the forward pass on typical activations.
Training asks for more: the operators must stay accurate on the wider range of values that fine-tuning produces (Section~\ref{sec:design:challenges}), return the values that their derivatives need, and still cost little communication.
VeilTune therefore designs its nonlinear operators for training.

VeilTune's operators rest on two ideas.
First, a polynomial approximation is accurate only on a small interval, so no operator approximates a function over its whole input range: inputs outside the interval are either computed by an exact formula, where one exists, or scaled into the interval by a power of two, which is local.
Second, the comparisons in these operators only decide which formula to apply or which value is larger, and a wrong decision near the boundary changes the result little, so they can run on truncated, low-precision copies of their inputs at less than half the cost.
Each operator also returns the values that its derivative needs.
Appendix~\ref{app:kernels} gives the coefficients and the error analysis of each operator.

\paragraph{GELU.}
GELU is almost exactly ReLU for $|\xi|\ge4$~\cite{sigma,bolt}, so VeilTune uses ReLU there and a polynomial only inside $[-4,4]$,
\begin{equation}
\mathrm{GELU}(\xi)\approx
\begin{cases}
\tfrac{\xi}{2}+R(\xi^2/16), & |\xi|<4,\\
\max(\xi,0), & |\xi|\ge 4,
\end{cases}
\label{eq:gelu-poly}
\end{equation}
where $R$ is a degree-7 polynomial, with a second polynomial for the derivative.
Which formula applies is decided by two low-precision comparisons, of $\xi$ with $0$ and of $\xi^2$ with $16$.

\paragraph{Softmax.}
Softmax subtracts the row maximum, exponentiates and normalizes.
The maximum is found by low-precision pairwise comparisons, since any value close to it serves equally well.
The exponential of the shifted scores, which are at most zero, is computed as $(E_6(a/128))^{128}$ for a shifted score $a$: dividing by 128 brings the input into a small interval where the degree-6 Taylor polynomial $E_6$ is accurate, and squaring seven times undoes the division.
Unlike prior designs~\cite{bumblebee,nimbus}, this needs no comparison to clip very small exponentials.

\paragraph{Inverse square root and reciprocal.}
LayerNorm and Adam need $v^{-1/2}$, and softmax needs $v^{-1}$, for $v$ from $10^{-4}$ to $10^{7}$.
VeilTune writes each argument as
\begin{equation}
v=u\cdot 2^{e}\ \text{ with }\ u\in[\tfrac12,1),
\label{eq:rangenorm}
\end{equation}
in the spirit of SIGMA~\cite{sigma}: the exponent $e$ is found by comparing $v$ with public powers of two, and dividing by $2^{e}$ is local.
Newton's iteration then only has to handle $u\in[\tfrac12,1)$, where two steps from one fixed starting point reach full precision, and scaling the result by $2^{-e/2}$ or $2^{-e}$ gives the answer for any $v$.

Together, these operators save 6.0\,GB and 457 rounds per step over the generic ones, and replacing them with the generic ones raises the relative error of the gradient to $8\times10^{8}$ (Section~\ref{sec:eval:ablation}).

\paragraph{Backward pass.}
With these operators, the backward pass consists of multiplications only.
At the logits, the gradient of the cross-entropy loss is $\partial\ell/\partial z=p-y$, which VeilTune obtains from the softmax output $p$ and the one-hot label $y$.
Through the base weights, the gradient is multiplied by a public matrix, which is local; through the nonlinear layers, it uses the values saved in the forward pass, for example
\begin{equation}
\delta_\xi=\delta_g\odot\mathrm{GELU}'(\xi),\qquad \delta_s=p\odot\bigl(\delta_p-\langle\delta_p,p\rangle\mathbf{1}\bigr)
\end{equation}
for GELU with output $g$ and softmax with output $p$, where $\delta_u$ denotes the gradient with respect to $u$, and LayerNorm works the same way with its saved normalized input and inverse standard deviation.
Finally, a LoRA branch that adds $hA_jB_j$ to a projection receives
\begin{equation}
\nabla_{B_j}=(hA_j)^{\top}\delta,\qquad \nabla_{A_j}=h^{\top}\bigl(\delta B_j^{\top}\bigr).
\end{equation}

\subsection{Oblivious Step Skipping}
\label{sec:design:adam}

Even with accurate operators, the fixed-point encoding has a limited range: large products cause truncation failures (Equation~\eqref{eq:truncfail}), and small values underflow to zero.
Both errors are most damaging in the optimizer, because Adam's moments $m$ and $v$ persist across steps: a truncation failure that reaches them distorts every later update, and a second moment that underflows to zero makes the step arbitrarily large.

Truncation failures are the harder of the two.
Local truncation~\cite{secureml} is what keeps secure training affordable, but a training step truncates millions of values, so some steps contain a failure.
Prior systems either accept this probability~\cite{secureml,crypten} or replace local truncation with exact truncation protocols at additional communication cost~\cite{sigma}.
Our insight is that training does not need every step: a failure shifts a value by about $2^{32}$, so the corrupted step is easy to recognize, and dropping it costs almost no accuracy (Section~\ref{sec:eval:acc}).
VeilTune therefore keeps local truncation and skips corrupted steps obliviously, so that neither party learns which steps were skipped.

\paragraph{Adam update.}
The backward pass delivers the adapter gradient $g$ in shares, and VeilTune updates the adapter with Adam (Equation~\eqref{eq:adam-std}).
The bias corrections depend only on the public step index $t$, so VeilTune folds them into the public constants $\eta_t$ and $\epsilon_t$ (Appendix~\ref{app:adam}) and, placing $\epsilon_t$ inside the square root, takes the step
\begin{equation}
\theta\leftarrow\theta-\eta_t\,q,\qquad q=\frac{m}{\sqrt{v+\epsilon_t^2}},
\label{eq:adam}
\end{equation}
with $m$ and $v$ updated as in Equation~\eqref{eq:adam-std}.

\paragraph{Skipping a step obliviously.}
A gradient computed from a truncation failure typically has entries of order $10^{7}$, whereas legitimate entries are of order one or smaller.
VeilTune compares every entry of $g$ with a public bound $\tau=10$, which separates the two magnitudes, and reduces the outcome to a single secret bit
\begin{equation}
o=\Bigl[\sum_j [\,|g_j|>\tau\,]=0\Bigr],
\end{equation}
which is $1$ exactly when no entry exceeds $\tau$; it computes the count with one batched comparison and compares it with zero.
Let $\omega'$ be the value that Equation~\eqref{eq:adam} and the moment updates would produce.
VeilTune then updates each state variable $\omega\in\{\theta,m,v\}$ as
\begin{equation}
\llbracket \omega\rrbracket\leftarrow\llbracket \omega\rrbracket+\llbracket o\rrbracket\cdot\bigl(\llbracket \omega'\rrbracket-\llbracket \omega\rrbracket\bigr).
\label{eq:skip}
\end{equation}
If $o=1$, the update is applied; if $o=0$, the adapter and both moments keep their previous values, so the corrupted gradient leaves no trace.
Since $o$ is never opened, neither party learns whether a step was skipped.

\paragraph{Scaling small values.}
Underflow, unlike truncation failures, affects almost every step, first in the gradient and then in the second moment.
Many entries of the adapter gradient lie below $2^{-f}\approx1.5\times10^{-5}$, the smallest magnitude the fixed-point encoding can represent, so VeilTune scales the gradient up by a public factor of 256 at the start of backpropagation and divides the adapter gradient by the same factor at the end~\cite{micikevicius2018mixed}; this lowers the relative error of the gradient from 16\% to 0.5\% (Section~\ref{sec:eval:ablation}).
The second moments are smaller still: when fine-tuning RoBERTa-large on SST-2, 91\% of them lie below $2^{-f}$, and so does $\epsilon^2$.
VeilTune therefore enlarges these values instead of representing them as they are.
It multiplies the gradient by a public constant $S=2^{13}$ before the Adam update, and $\epsilon_t$ by the same constant.
The first moment then becomes $Sm$ and the second moment $S^2v$, which is $2^{26}$ times larger and thus representable down to $2^{-42}\approx2\times10^{-13}$, while the step is unchanged, because the scale cancels:
\begin{equation}
\frac{Sm}{\sqrt{S^2v+S^2\epsilon_t^2}}=\frac{m}{\sqrt{v+\epsilon_t^2}}=q.
\end{equation}
The scaled moments must not grow without bound, however.
VeilTune clips every gradient entry that passes the check above to $[-1,1]$ on shares, so that $|Sm|\le S$ and $0\le S^2v\le S^2$.
A truncation failure inside the update itself, which the check above cannot see, is caught by a range check on the moments (Appendix~\ref{app:adam}).

\subsection{Security}
\label{sec:design:security}

Multiplication and comparison on shares are modeled by ideal functionalities $\mathcal{F}_{\textsf{Mul}}$ and $\mathcal{F}_{\textsf{Cmp}}$, realized with correlated randomness from the preprocessing phase (Section~\ref{sec:setting:threat}).
On inputs $\llbracket a\rrbracket,\llbracket a'\rrbracket$, $\mathcal{F}_{\textsf{Mul}}$ returns fresh shares of the truncated product $aa'$, and $\mathcal{F}_{\textsf{Cmp}}$ returns fresh shares of $[a<0]$; each output share is uniform over its ring and independent of all other values.

\begin{theorem}
\label{thm:security}
In the $(\mathcal{F}_{\textsf{Mul}},\mathcal{F}_{\textsf{Cmp}})$-hybrid model, VeilTune satisfies Definition~\ref{def:security}.
\end{theorem}

\begin{proof}
After both parties secret-share their inputs, write the whole training run as a sequence of $M$ calls $c_1,\dots,c_M$ to $\mathcal{F}_{\textsf{Mul}}$ or $\mathcal{F}_{\textsf{Cmp}}$, interleaved with local operations, followed by the opening of $\theta_T$ to $P_1$.
Let $\sigma_m=(\mathcal{F}^{(m)},\nu_m,k_m)$ be the functionality of call $c_m$, the number of entries and the bit width of its output.
The schedule $\sigma=(\sigma_1,\dots,\sigma_M)$ is a deterministic function of $\mathcal{L}(\mathcal{D})$: token packing and the \texttt{[MASK]}-only rows are fixed by the lengths and \texttt{[MASK]} positions, and skipping selects the update through multiplication with the secret bit $o$ in Equation~\eqref{eq:skip}.
Local operations are deterministic functions of a party's own shares, so in the hybrid model the view of $P_i$ is
\begin{equation*}
\mathsf{view}_i=\bigl(x_i,\ \llbracket x_{1-i}\rrbracket_i,\ \llbracket \zeta_1\rrbracket_i,\dots,\llbracket \zeta_M\rrbracket_i,\ \mu_i\bigr),
\end{equation*}
where $x_0=\mathcal{D}$ and $x_1=\theta_0$ are the inputs, $\llbracket x_{1-i}\rrbracket_i$ is the share of the other party's input, $\llbracket \zeta_m\rrbracket_i\in\mathbb{Z}_{2^{k_m}}^{\nu_m}$ is the output share of $c_m$, and $\mu_i$ is the message received in the final opening, with $\mu_0$ empty and $\mu_1=\llbracket\theta_T\rrbracket_0$.

\emph{Corrupted $P_0$.}
$\mathcal{S}_0(\mathcal{D},\mathcal{L})$ computes $\sigma$ from $\mathcal{L}$, samples a uniform share $\rho_0$ of an input of the shape of $\theta_0$ and $\rho_m\leftarrow\mathbb{Z}_{2^{k_m}}^{\nu_m}$ uniformly and independently for $m=1,\dots,M$, and outputs $(\mathcal{D},\rho_0,\dots,\rho_M)$.
Since $\llbracket\theta_0\rrbracket_0$ and each $\llbracket \zeta_m\rrbracket_0$ are uniform and independent of $\mathcal{D}$, $\theta_0$ and all other shares, $\mathcal{S}_0(\mathcal{D},\mathcal{L})\equiv\mathsf{view}_0$; truncation failures affect only the values $\zeta_m$, which never appear in the view.

\emph{Corrupted $P_1$.}
$\mathcal{S}_1(\theta_0,\theta_T,\mathcal{L})$ samples a uniform share $\rho_0$ of an input of the shape given by $\mathcal{L}$ and $\rho_1,\dots,\rho_M$ as above, computes from them by the local operations of $\Pi$ the share $\llbracket\theta_T\rrbracket_1$ that $P_1$ would hold, and sets $\mu_1=\theta_T\cdot2^{f}-\llbracket\theta_T\rrbracket_1 \bmod 2^{64}$.
The shares $\rho_m$ are distributed as in the real execution for the same reason as above, and given them, $\mu_1$ is the unique value that reconstructs the output $\theta_T$, so $\mathcal{S}_1(\theta_0,\theta_T,\mathcal{L})\equiv\mathsf{view}_1$.

Both simulations are perfect, which implies the computational indistinguishability of Definition~\ref{def:security}; replacing $\mathcal{F}_{\textsf{Mul}}$ and $\mathcal{F}_{\textsf{Cmp}}$ by their semi-honest realizations preserves security by the composition theorem~\cite{canetti}.
\end{proof}

\section{Evaluation}
\label{sec:eval}

\begin{table*}[t]
\caption{Cost of one secure training step by stage, per party, on the four tasks (batch size 32). Time is in seconds on a LAN (3\,Gbps, 0.8\,ms) and a WAN (200\,Mbps, 40\,ms); communication is in GB. The offline phase generates the correlated randomness and is independent of the inputs. The last row gives a full training run (150 steps; 100 on RTE) in hours and TB.}
\label{tab:cost}
\footnotesize\setlength{\tabcolsep}{3.2pt}
\begin{tabularx}{\textwidth}{@{}ll*{12}{R}@{}}
\toprule
 & & \multicolumn{3}{c}{SST-2} & \multicolumn{3}{c}{Yelp} & \multicolumn{3}{c}{AG News} & \multicolumn{3}{c}{RTE} \\
\cmidrule(lr){3-5}\cmidrule(lr){6-8}\cmidrule(lr){9-11}\cmidrule(l){12-14}
\multicolumn{2}{@{}l}{Stage} & LAN & WAN & Comm. & LAN & WAN & Comm. & LAN & WAN & Comm. & LAN & WAN & Comm. \\
\midrule
Offline & Correlated randomness & -- & -- & 6.8 & -- & -- & 36.4 & -- & -- & 19.0 & -- & -- & 30.4 \\
\midrule
\multirow{5}{*}{Forward} & Linear layers & 4.9 & 34.6 & 0.7 & 22.1 & 132 & 3.2 & 12.3 & 75.0 & 1.8 & 18.2 & 89.9 & 3.1 \\
 & Softmax & 8.4 & 99.7 & 3.1 & 57.9 & 660 & 30.4 & 24.8 & 295 & 12.6 & 55.8 & 658 & 30.4 \\
 & GELU & 26.5 & 322 & 15.3 & 139 & 1{,}644 & 79.8 & 72.5 & 901 & 43.7 & 102 & 1{,}231 & 60.1 \\
 & LayerNorm & 3.8 & 39.9 & 0.6 & 9.5 & 91.0 & 3.0 & 5.7 & 61.9 & 1.7 & 12.7 & 66.6 & 2.3 \\
 & Loss & 0.1 & 1.7 & 0.0 & 0.1 & 1.8 & 0.0 & 0.1 & 2.6 & 0.0 & 0.1 & 1.3 & 0.0 \\
\cmidrule(l){2-14}
Backward & & 9.5 & 89.9 & 2.9 & 47.3 & 424 & 17.1 & 24.3 & 222 & 8.6 & 39.2 & 345 & 15.1 \\
Update & Adam and skipping & 6.3 & 76.0 & 3.3 & 6.8 & 76.3 & 3.3 & 6.4 & 76.3 & 3.3 & 6.3 & 73.2 & 3.3 \\
\midrule
\multicolumn{2}{@{}l}{\textbf{Online step}} & \textbf{59.6} & \textbf{664} & \textbf{25.9} & \textbf{283} & \textbf{3{,}029} & \textbf{136.9} & \textbf{146} & \textbf{1{,}634} & \textbf{71.5} & \textbf{234} & \textbf{2{,}464} & \textbf{114.2} \\
\multicolumn{2}{@{}l}{Training run (h, TB)} & 2.5 & 27.7 & 3.9 & 11.8 & 126.2 & 20.5 & 6.1 & 68.1 & 10.7 & 6.5 & 68.4 & 11.4 \\
\bottomrule
\end{tabularx}
\end{table*}

\subsection{Setup}
\label{sec:eval:setup}

\paragraph{Model, tasks and configuration.}
We fine-tune RoBERTa-large~\cite{roberta} with the prompt-based classifier of LM-BFF~\cite{lmbff} on SST-2~\cite{sst2}, Yelp Polarity~\cite{zhang2015}, AG News~\cite{zhang2015} and RTE~\cite{rte}, with 16 training examples per class.
We report test accuracy, averaged over three seeds that each draw their own training examples.
The adapter has rank 4 on the query and value projections of the last eight blocks, and is trained with Adam at batch size 32; Appendix~\ref{app:setup} lists the remaining hyperparameters.

\paragraph{Implementation.}
VeilTune is built on CrypTen~\cite{crypten} with sharing over $\mathbb{Z}_{2^{64}}$ and $f=16$, and correlated randomness from a trusted dealer; we report the online phase.
We run the two parties on two servers, each equipped with an A100 GPU, and use Linux Traffic Control to emulate the network between them: 3\,Gbps with 0.8\,ms round-trip latency for LAN, and 200\,Mbps with 40\,ms round-trip latency for WAN.

\paragraph{Baselines.}
We compare with SecP-Tuning~\cite{secptuning} and with full-parameter fine-tuning in MPC (SFT); since our configuration follows that of SecP-Tuning, we take the results of both from~\cite{secptuning}.
To separate the effect of the optimizer from that of the adapter, we also train our adapter in plaintext with CMA-ES, the optimizer of SecP-Tuning, and with two-point SPSA~\cite{spall1992spsa,malladi2023mezo} for the forward-only comparison.
For the cost of a step, the baseline is the same step built from generic operators, which we instantiate with CrypTen's default implementations.

\subsection{Accuracy}
\label{sec:eval:acc}

\begin{table}[t]
\caption{Few-shot test accuracy (\%, mean $\pm$ standard deviation over three seeds; higher is better). SFT and SecP-Tuning are reported in~\cite{secptuning}; the other rows train our adapter, in plaintext except for the last row; LoRA + SPSA is zeroth-order and given as a mean. The best result in each column is in bold and the second best is underlined.}
\label{tab:acc}
\footnotesize
\setlength{\tabcolsep}{1.5pt}
\begin{tabularx}{\columnwidth}{@{}l*{5}{C}@{}}
\toprule
\multirow{2}{*}{Method} & \multicolumn{5}{c}{Test accuracy (\%) $\uparrow$} \\
\cmidrule(l){2-6}
 & SST-2 & Yelp & AG News & RTE & Avg. \\
\midrule
SFT & 85.4\,$\pm$\,2.8 & \underline{91.8}\,$\pm$\,0.8 & 86.4\,$\pm$\,1.9 & 58.6\,$\pm$\,6.2 & 80.5 \\
SecP-Tuning & \textbf{88.1}\,$\pm$\,0.6 & 85.2\,$\pm$\,0.5 & 81.3\,$\pm$\,0.7 & 53.0\,$\pm$\,0.3 & 76.9 \\
\midrule
LoRA + CMA-ES & 85.1\,$\pm$\,3.4 & 91.1\,$\pm$\,1.4 & 86.6\,$\pm$\,2.0 & 63.9\,$\pm$\,2.9 & 81.7 \\
LoRA + SPSA & 85.6 & 91.1 & 86.0 & 64.9 & 81.9 \\
LoRA + Adam & \underline{86.7}\,$\pm$\,1.1 & \textbf{92.0}\,$\pm$\,0.9 & \textbf{87.6}\,$\pm$\,1.0 & \textbf{66.5}\,$\pm$\,1.7 & \textbf{83.2} \\
\midrule
\rowcolor{gray!15}VeilTune & 86.4\,$\pm$\,1.3 & 90.3\,$\pm$\,1.8 & \underline{87.0}\,$\pm$\,1.3 & \underline{65.0}\,$\pm$\,0.7 & \underline{82.2} \\
\bottomrule
\end{tabularx}
\end{table}

\paragraph{Backpropagation gives a more accurate adapter.}
Table~\ref{tab:acc} compares the training algorithms in plaintext.
The algorithm that VeilTune runs on shares, LoRA trained with Adam, is 6.3 points above SecP-Tuning on average, with the largest gain, 13.5 points, on RTE; only on SST-2 does SecP-Tuning report 1.4 points more.
It is also more accurate than SFT on every task.
Training the same adapter with CMA-ES splits the gap to SecP-Tuning: 4.8 points come from tuning an adapter instead of a prompt, and 1.5 from backpropagation; the zeroth-order optimizer SPSA~\cite{spall1992spsa,malladi2023mezo}, which updates the adapter from the losses of two forward passes, also falls behind Adam on every task.

\paragraph{Secure training stays close to plaintext training.}
The last two rows of Table~\ref{tab:acc} train the same adapter on the same examples in plaintext and in MPC.
VeilTune is 0.3 points below plaintext training on SST-2, with differences of both signs across seeds, and 0.6, 1.5 and 1.7 points below on AG News, RTE and Yelp.
The gap comes from the skipped steps, whose share grows with the input length, since longer inputs contain more products that can fail: about 10\% on SST-2, 40\% on AG News and RTE, and half on Yelp, whose inputs average 17, 56, 68 and 100 tokens.
Skipping them costs little, since plaintext training with and without these steps differs by less than 0.3 points on every task, whereas keeping them is fatal: without skipping, the 38 corrupted steps of an RTE run enter the update and the accuracy stays near chance, at 52.7\% against 65.7\%.

\subsection{Training Cost}
\label{sec:eval:cost}\label{sec:eval:zo}

\paragraph{The nonlinear operators of the forward pass dominate the cost.}
Table~\ref{tab:cost} breaks a training step into its stages on all four tasks.
On every task, GELU and softmax account for most of the communication and time, because they evaluate the nonlinear operators of eight blocks on every token.
The backward pass adds only about one seventh of the forward pass, since it multiplies by public weights and by values saved in the forward pass (Section~\ref{sec:design:backward}), and the update costs the same 3.3\,GB on every task, since it runs on the adapter entries rather than on the tokens.
The offline phase adds about a quarter of the online communication and can run before training starts.

\paragraph{VeilTune costs less than SecP-Tuning end to end.}
The last row of Table~\ref{tab:cost} gives a full training run.
Counting one parameter update as a step, VeilTune converges within 150 steps of one forward and one backward pass, whereas SecP-Tuning evaluates 20 candidate prompts per step for up to 400 steps~\cite{secptuning}.
On SST-2 under the same LAN setting, VeilTune takes 2.5 hours and 3.9\,TB of communication per party, against 8.8 hours and 14.2\,TB for SecP-Tuning and 65.9 hours and 67.4\,TB for SFT, as reported on their own hardware~\cite{secptuning}.
On AG News, whose inputs are about three times longer, VeilTune communicates 10.7\,TB, against 19.7 and 121.3\,TB.
Forward-only training would not be cheaper either: SPSA needs 929 steps to come within one point of its final accuracy, against 58 for Adam, and each of its steps evaluates the forward pass twice, which costs 39.4\,GB on SST-2 against 25.9\,GB for a VeilTune step, so it needs about 24 times more communication to converge.

\subsection{Ablation}
\label{sec:eval:ablation}

\begin{table}[t]
\caption{Effect of removing one technique at a time (one training step on SST-2 without the optimizer, batch size 32, per party). The last column is the relative error of the adapter gradient against float64.}
\label{tab:ablation}
\footnotesize
\setlength{\tabcolsep}{4pt}
\begin{tabularx}{\columnwidth}{@{}l*{3}{R}@{}}
\toprule
Variant & GB $\downarrow$ & Rounds $\downarrow$ & Grad.\ error $\downarrow$ \\
\midrule
\rowcolor{gray!15}\textbf{VeilTune} & 22.6 & 1{,}208 & 0.54\% \\
\midrule
w/o token packing & 42.0 & 1{,}208 & -- \\
w/o \texttt{[MASK]}-only rows & 25.5 & 1{,}208 & -- \\
w/o loss scaling & -- & -- & 15.7\% \\
generic operators & 28.6 & 1{,}665 & $8.0\times10^{8}$ \\
\midrule
all removed & 55.3 & 1{,}665 & $2.0\times10^{11}$ \\
\bottomrule
\end{tabularx}
\end{table}

\paragraph{Each technique pays for itself.}
Table~\ref{tab:ablation} removes one technique at a time.
Token packing saves the most communication: without it, a step needs 42.0 instead of 22.6\,GB, and computing every row of the last block adds another 2.9\,GB.
The training-aware operators save 457 rounds and 6.0\,GB, and without them the relative error of the gradient rises from 0.54\% to $8\times10^{8}$; without loss scaling, it rises to 15.7\%.
With all techniques removed, a step needs 2.4 times the communication of VeilTune.

\section{Conclusion}
\label{sec:concl}

We presented VeilTune, which runs the whole training step in MPC and reveals only the trained adapter, to the provider alone.
With token packing, training-aware nonlinear operators and oblivious step skipping, it nearly matches plaintext accuracy and is both more accurate and cheaper than SecP-Tuning.

\begin{acks}
This work was supported in part by the National Key Research and Development Program of China (No.~2023YFB2906400;\newline No.~2024YFB2907202); the Natural Science Basic Research Program of Shaanxi under Grant No.~2026JC-YBMS-0669; the National Cryptologic Science Fund of China under No.~2025NCSF01007; the National Natural Science Foundation of China Key Program under Grant No.~92270204; and the CAS Project for Young Scientists in Basic Research under Grant No.~YSBR-035.
\end{acks}
\label{lastbodypage}
\bibliographystyle{ACM-Reference-Format}
\bibliography{refs}
\appendix
\section{Detailed Leakage Results}
\label{app:leak}

The adapter has rank 4 on the query and value projections of the last eight blocks; before the protocol starts, the model owner trains it on 4,000 held-out examples (on RTE, whose training set is small, on the 1,905 examples of the pool that no data owner draws), which brings it to 91.2\% accuracy on SST-2, and the protocol then fine-tunes it for 1,000 steps on the data owner's 16 labeled examples per class.
SAP and PrivTuner update the adapter with AdamW at learning rate $10^{-4}$ and mini-batches of 32 examples.
For SecP-Tuning, CMA-ES~\cite{hansen2016cma} with population 20 searches a random 500-dimensional subspace of the LoRA parameters, following black-box tuning~\cite{bbt}, so that all three protocols protect the same asset, and the data owner observes the logits of every candidate.
The active attacker adds 512 unlabeled public sentences to each of its mini-batches.
The extracted adapter $\hat\theta$ is trained with Adam at learning rate $10^{-3}$ for 3,000 iterations, and the local model for 300 steps on the data owner's examples.

All results are averaged over three seeds.
Table~\ref{tab:attack} gives the test accuracy and the agreement with $\theta_T$ behind Figure~\ref{fig:leak}.
On every task, $\hat\theta$ extracted from hidden representations, or from logits by the active attacker, is close to $\theta_T$ in accuracy, so the gain in agreement does not come at the cost of accuracy.

\begin{table}[h]
\caption{Adapter leakage: test accuracy / agreement with $\theta_T$ (\%) of the adapter $\hat\theta$ extracted by AEA from the last $K$ steps the protocol opens; rows without ``+ active'' are passive attackers. The second column gives the test accuracy of $\theta_T$, and the third the agreement of the local model.}
\label{tab:attack}
\footnotesize
\setlength{\tabcolsep}{3.5pt}
\begin{tabularx}{\columnwidth}{@{}l*{4}{C}@{}}
\toprule
Protocol (opened values) & $\theta_T$ acc. & Local & $K=1$ & $K=100$ \\
\midrule
\multicolumn{5}{@{}l}{\emph{SST-2} (local model accuracy 86.4\%)} \\
SAP (hidden repr.) & 90.4 & 90.6 & 90.6 / 99.2 & 90.6 / 99.1 \\
PrivTuner (logits) & 90.4 & 90.6 & 87.4 / 93.2 & 87.4 / 92.7 \\
\quad + active & 90.4 & 90.6 & 90.9 / 97.8 & 90.9 / 97.7 \\
SecP-Tuning (logits) & 87.5 & 88.1 & 86.7 / 91.0 & 88.0 / 90.6 \\
\quad + active & 87.5 & 88.1 & 87.9 / 94.3 & 87.4 / 94.5 \\
\midrule
\multicolumn{5}{@{}l}{\emph{Yelp} (local model accuracy 90.8\%)} \\
SAP (hidden repr.) & 95.0 & 92.2 & 94.8 / 99.5 & 94.8 / 99.5 \\
PrivTuner (logits) & 95.0 & 92.2 & 92.2 / 93.8 & 92.3 / 94.1 \\
\quad + active & 95.0 & 92.2 & 94.7 / 98.3 & 94.8 / 98.4 \\
SecP-Tuning (logits) & 92.5 & 90.2 & 92.4 / 92.0 & 92.7 / 92.2 \\
\quad + active & 92.5 & 90.2 & 93.7 / 94.7 & 93.8 / 94.7 \\
\midrule
\multicolumn{5}{@{}l}{\emph{AG News} (local model accuracy 86.8\%)} \\
SAP (hidden repr.) & 90.6 & 89.8 & 90.4 / 98.6 & 90.5 / 99.3 \\
PrivTuner (logits) & 90.6 & 89.8 & 87.0 / 90.0 & 88.3 / 92.0 \\
\quad + active & 90.6 & 89.8 & 89.9 / 95.3 & 90.1 / 95.6 \\
SecP-Tuning (logits) & 89.1 & 89.8 & 87.0 / 89.7 & 88.5 / 92.2 \\
\quad + active & 89.1 & 89.8 & 89.2 / 93.9 & 89.6 / 94.2 \\
\midrule
\multicolumn{5}{@{}l}{\emph{RTE} (local model accuracy 65.1\%)} \\
SAP (hidden repr.) & 76.8 & 72.4 & 74.8 / 93.0 & 75.0 / 92.7 \\
PrivTuner (logits) & 76.8 & 72.4 & 62.0 / 66.7 & 61.5 / 66.2 \\
\quad + active & 76.8 & 72.4 & 75.0 / 87.8 & 75.1 / 86.8 \\
SecP-Tuning (logits) & 72.7 & 72.0 & 62.5 / 69.1 & 61.9 / 69.4 \\
\quad + active & 72.7 & 72.0 & 72.7 / 82.9 & 72.4 / 82.9 \\
\bottomrule
\end{tabularx}
\end{table}

\section{Details of the Nonlinear Operators}
\label{app:kernels}

\begin{figure}[h]
\centering
\includegraphics[width=\columnwidth]{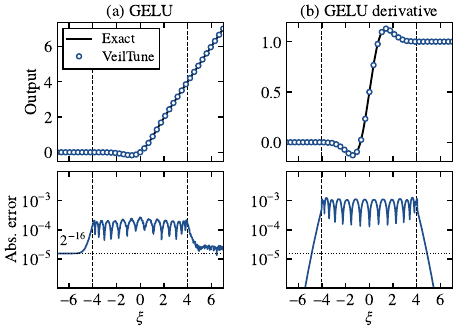}
\caption{GELU and its derivative as computed by VeilTune in 16-bit fixed point, against the exact erf-based GELU. The polynomial is used between the dashed lines; the bottom row shows the absolute error.}
\label{fig:approx}
\end{figure}

\paragraph{GELU.}
RoBERTa uses the exact GELU, $\mathrm{GELU}(\xi)=\xi\,\Phi(\xi)$, where $\Phi$ is the standard normal distribution function.
We write it as $\mathrm{GELU}(\xi)=\xi/2+H(\xi)$ with $H(\xi)=\xi\,(\Phi(\xi)-\tfrac12)$.
The function $H$ is even and approaches $|\xi|/2$ quickly, so GELU coincides with ReLU up to $1.3\times10^{-4}$ for $|\xi|\ge 4$.
On $|\xi|<4$ we approximate $H$ and the derivative by polynomials in $\upsilon=\xi^2/16\in[0,1)$,
\begin{equation}
H(\xi)\approx R(\upsilon),\qquad \mathrm{GELU}'(\xi)-\tfrac12\approx \xi\,Q(\upsilon),
\end{equation}
where $R$ has degree 7 and $Q$ degree 6, both fitted with the minimax criterion.
Writing $\chi=[\xi^2<16]$ and $\psi=1-2[\xi<0]$, the operator outputs
\begin{align}
\mathrm{GELU}(\xi) &\approx \tfrac{\xi}{2}+\tfrac{\psi \xi}{2}+\chi\,\bigl(R(\upsilon)-\tfrac{\psi \xi}{2}\bigr),\\
\mathrm{GELU}'(\xi) &\approx \tfrac{1}{2}+\tfrac{\psi}{2}+\chi\,\bigl(\xi\,Q(\upsilon)-\tfrac{\psi}{2}\bigr).
\end{align}
The two bits come from one batched comparison.
$R$ and $Q$ are evaluated with Horner's rule side by side, so that each Horner step of both polynomials is one batched multiplication.
With the branch bits computed on truncated copies (see below), the maximum absolute error in 16-bit fixed point is $2.7\times10^{-4}$ for GELU and $1.3\times10^{-3}$ for its derivative (Figure~\ref{fig:approx}).

\paragraph{Softmax.}
For a row of scores $\alpha\in\mathbb{R}^{n'}$, the pairwise reduction replaces each pair $a,a'$ by $\max(a,a')=a+[a-a'<0]\cdot(a'-a)$, which costs one comparison and one multiplication, so a row needs $\lceil\log_2 n'\rceil$ levels.
The exponential of $a=\alpha_i-\max_{i'}\alpha_{i'}\le 0$ is computed as
\begin{equation}
e^{a}\approx \bigl(E_6(a/2^{7})\bigr)^{2^{7}},\qquad E_6(w)=\sum_{i=0}^{6}\frac{w^i}{i!}.
\end{equation}
A Taylor polynomial of even degree of $e^w$ is positive on the whole real line, and $E_6(w)<1$ for $w\in(-3.55,0)$, so the approximation stays in $(0,1)$ for every $a\in(-454,0]$; observed attention scores lie well inside this range.
Prior designs instead set the exponential to zero below a threshold, which costs a comparison~\cite{bumblebee,nimbus}.

\paragraph{Inverse square root.}
A single batched comparison against the public ladder $2^{i}$, $i\in\{-14,\dots,26\}$, yields bits $[v<2^{i}]$, and the differences of consecutive bits form a one-hot vector that marks the exponent $e$ with $u=v\cdot 2^{-e}\in[\tfrac12,1)$.
Since scaling by a public power of two is local, $u$ is obtained from the one-hot vector with one batched multiplication.
On $[\tfrac12,1)$ the linear guess $w_0=c_0-c_1u$ with public $c_0,c_1$ approximates $u^{-1/2}$ within 2.3\%, and two Newton steps $w_{i+1}=\tfrac12\,w_i\,(3-u\,w_i^2)$ reduce the relative error to below $10^{-6}$ in exact arithmetic, that is, to the fixed-point resolution.
The result is $v^{-1/2}=w_2\cdot 2^{-e/2}$, where an odd $e$ contributes a public factor $\sqrt{1/2}$, and is selected with one more multiplication by the one-hot vector.
The cost is one comparison and six rounds of multiplication, independent of the range.
The reciprocal of the softmax row sum, which lies in $[1,n']$, uses the same normalization with $\lceil\log_2 n'\rceil$ thresholds and three Newton steps.

\paragraph{LayerNorm.}
With the public constant $\lambda=\tfrac14$ and the centered input $a\in\mathbb{R}^{d}$,
\begin{equation}
\frac{1}{\sqrt{\mathrm{Var}(a)}}=\lambda\cdot\Bigl(\tfrac{1}{d}\textstyle\sum_{i=1}^{d}(\lambda\,a_i)^2\Bigr)^{-1/2},
\end{equation}
so the squared products shrink by $1/\lambda^2=16$ and so does their failure probability by Equation~\eqref{eq:truncfail}.
In the upper blocks of RoBERTa the centered input reaches $|a_i|\approx1.5\times10^{3}$ in outlier dimensions.

\paragraph{LayerNorm backward.}
With the saved normalized input $\hat a$, inverse standard deviation $\varrho$ and public scale $\gamma$, the gradient with respect to the input is $\delta_a=\varrho\,(\delta_{\hat a}-\overline{\delta_{\hat a}}-\hat a\,\overline{\delta_{\hat a}\odot\hat a})$ with $\delta_{\hat a}=\gamma\odot\delta$, where $\delta$ is the gradient with respect to the output and the bar denotes the mean over the hidden dimension.

\paragraph{Comparisons on truncated values.}
The branch bits of GELU and the pairwise comparisons for the softmax maximum are computed on truncated values.
To compute $[a<0]$ for a shared value $a$ known to satisfy $|a|<2^{\kappa+k-1}$ in ring units, each party truncates its share by $\kappa$ bits locally, as in SecureML~\cite{secureml}, and reduces the result modulo $2^{k}$, which is also local because reduction modulo a power of two commutes with addition.
The two reduced shares sum to $\lfloor a/2^{\kappa}\rceil$ modulo $2^{k}$, up to an error of one unit, so their sum has the sign bit of $a$ whenever $|a|\ge 2^{\kappa}$.
Each party packs $64/k$ reduced shares into the lanes of one 64-bit word, and the parties add the packed words with CrypTen's binary adder restricted to its first $\log_2k$ levels, whose masks are periodic in $k$ bits, so that no carry crosses a lane boundary.
The most significant bit of each lane is the comparison result and is converted to an arithmetic share by the usual single-bit conversion.
For GELU we use $\kappa=16$ and $k=32$, which is correct for $|\xi|$ up to $4.6\times10^{4}$ and misjudges only $|\xi|<1$ for the sign and $|\xi^2-16|<1$ for the interval, where both branches agree; for the softmax maximum we use $\kappa=12$ and $k=16$, which is correct for score differences up to 2,048 and errs only for differences below $1/16$.
A comparison on the full 64-bit value costs 432 bytes per element and party; on truncated values it costs 192 bytes for $k=32$ and 88 bytes for $k=16$.

\section{Details of Secure Adam}
\label{app:adam}

\paragraph{Constants.}
We use $\beta_1=0.9$, $\beta_2=0.999$ and $\epsilon=3\times10^{-5}$.
The bias corrections of Adam~\cite{kingma2015adam} are folded into the public constants
\begin{equation}
\eta_t=\eta\,\frac{\sqrt{1-\beta_2^{t}}}{1-\beta_1^{t}},\qquad \epsilon_t^2=(1-\beta_2^{t})\,\epsilon^2,
\end{equation}
so that Equation~\eqref{eq:adam} with $m$ and $v$ initialized to zero equals the bias-corrected update of Equation~\eqref{eq:adam-std} with $\epsilon$ placed inside the square root.
In the scaled domain $\epsilon_t$ is replaced by $S\epsilon_t$.

\paragraph{Second moment.}
The scaled second moment is accumulated as $v\leftarrow\beta_2v+(\alpha\,Sg)^2$ with the public constant $\alpha=\sqrt{1-\beta_2}$, so that the product to be truncated is $\alpha^2(Sg)^2$ rather than $(Sg)^2$; it is $1/\alpha^2=1000$ times smaller, and so is its failure probability (Equation~\eqref{eq:truncfail}).

\paragraph{Range checks on the moments.}
The legal ranges of Section~\ref{sec:design:adam} are widened by 1\% to absorb fixed-point rounding, that is, $|Sm|\le1.01\,S$ and $-1\le S^2v\le1.01\,S^2$.
All entries of both moments are checked in one batched comparison, and the per-entry bits are multiplied by the step bit $o$ before Equation~\eqref{eq:skip} is applied.

\paragraph{Inverse square root.}
The inverse square root of Equation~\eqref{eq:adam} is element-wise, so it is evaluated once on the concatenation of all adapter tensors; its comparison and its Newton rounds are thus paid once per step, independent of the number of adapted layers.

\paragraph{Clipping the step.}
As a last safeguard, the normalized step $q$ is clipped to $[-4,4]$ before the update.
A legitimate step never reaches this bound, since $|q|\le(1-\beta_1)/\sqrt{1-\beta_2}\approx3.2$~\cite{kingma2015adam}, so the clipping affects only entries that a residual numerical error has pushed out of range.

\section{Experimental Details}
\label{app:setup}

\paragraph{Data.}
Inputs are truncated to 64 tokens on SST-2 and to 128 tokens on the other tasks; the average input has 17 tokens on SST-2, 56 on AG News, 68 on RTE and 100 on Yelp.
Each seed draws its own 16 training examples per class, and checkpoints of plaintext runs are selected on a held-out development set.

\paragraph{VeilTune.}
Adam runs for 150 steps (100 on RTE) with learning rate $10^{-4}$ ($3\times10^{-4}$ on AG News and RTE), loss scale 256, optimizer scale $S=2^{13}$ and skipping bound $\tau=10$; Appendix~\ref{app:adam} gives the constants of the optimizer and Appendix~\ref{app:hparam} the sensitivity to these choices.

\paragraph{Baselines.}
LoRA + CMA-ES follows black-box tuning~\cite{bbt}, on which SecP-Tuning builds: it searches a random 500-dimensional subspace of the adapter with CMA-ES~\cite{hansen2016cma}, with population 20, a budget of 8,000 forward evaluations, and the same templates and label words as VeilTune.
SPSA uses perturbation $10^{-3}$ and learning rates in $\{10^{-4},3\times10^{-4},10^{-3}\}$ selected per seed, for up to 4,000 steps.

\section{Hyperparameter Sensitivity}
\label{app:hparam}

VeilTune adds three hyperparameters to plaintext training: the loss scale, the optimizer scale $S$ and the skipping bound $\tau$.
The learning rate is chosen in plaintext from $\{10^{-4},3\times10^{-4}\}$ on the development set and reused in MPC.

\paragraph{Loss scale.}
Table~\ref{tab:hp-ls} gives the relative error of the secure gradient with respect to the plaintext gradient for one step on three batches of four examples; Table~\ref{tab:ablation} measures the full training batch, on which the error without scaling is 15.7\%.
Without scaling, small gradient entries fall below the resolution of the fixed-point encoding and the error reaches 19\%; from a scale of 16 on, it stays below 1.1\%, and it is flat between 64 and 4,096.
The default of 256 lies in the middle of this range.

\begin{table}[h]
\caption{Relative error (\%) of the secure gradient for different loss scales: median and range over three batches of SST-2.}
\label{tab:hp-ls}
\footnotesize
\begin{tabularx}{\columnwidth}{@{}l*{6}{C}@{}}
\toprule
Loss scale & 1 & 16 & 64 & 256 & 1{,}024 & 4{,}096 \\
\midrule
Median & 13.3 & 1.03 & 0.59 & 0.48 & 0.62 & 0.60 \\
Range & 3.5--19.1 & 0.33--1.08 & 0.24--0.60 & 0.22--0.71 & 0.21--0.64 & 0.20--0.60 \\
\bottomrule
\end{tabularx}
\end{table}

\paragraph{Skipping bound.}
The bound $\tau$ separates steps corrupted by a truncation failure from normal ones by the largest gradient entry.
Over the 1,650 steps of the twelve secure runs in Section~\ref{sec:eval:acc}, this entry is below 1 in 1,037 steps and above $10^{6}$ in 532, with only 81 steps in between and none between $10^{4}$ and $10^{6}$.
Any $\tau$ from 1 to $10^{6}$ therefore skips nearly the same steps: on SST-2 the share of skipped steps only falls from 11.6\% at $\tau=1$ to 9.3\% at $\tau=10^{4}$, and on the other tasks it changes by at most 7.6 points.

\paragraph{End-to-end training.}
Table~\ref{tab:hp-run} changes one hyperparameter at a time and trains on SST-2 in MPC.
Because the plaintext reference skips the same steps as the secure run, its accuracy varies slightly with the skipped steps; the secure run stays within 0.8 points of it in every setting.

\begin{table}[h]
\caption{Secure training on SST-2 (seed 42) with one hyperparameter changed at a time: test accuracy (\%) of the secure run and of plaintext training that skips the same steps, and the number of skipped steps out of 150.}
\label{tab:hp-run}
\footnotesize
\begin{tabularx}{\columnwidth}{@{}l*{3}{C}@{}}
\toprule
Setting & Secure & Plaintext & Skipped \\
\midrule
\rowcolor{gray!15}Default ($\tau=10$, scale 256, $S=2^{13}$) & 87.6 & 87.7 & 18 \\
$\tau=1$ & 87.8 & 87.7 & 13 \\
$\tau=1{,}000$ & 87.3 & 87.7 & 12 \\
Loss scale 64 & 88.9 & 88.8 & 28 \\
Loss scale 1{,}024 & 89.0 & 88.7 & 19 \\
$S=2^{11}$ & 87.6 & 87.7 & 13 \\
$S=2^{15}$ & 87.2 & 87.7 & 17 \\
\bottomrule
\end{tabularx}
\end{table}

\section{Notation}
\label{app:notation}

\begin{table}[H]
\caption{Notation.}
\label{tab:notation}
\footnotesize
\begin{tabularx}{\columnwidth}{@{}lX@{}}
\toprule
Symbol & Meaning \\
\midrule
$P_0$, $P_1$ & data owner, model owner \\
$\llbracket x\rrbracket$ & additive sharing of $x$ over $\mathbb{Z}_{2^{64}}$ \\
$f$ & fractional bits of the fixed-point encoding ($f=16$) \\
$(x,y)\in\mathcal{D}$ & private input and label of $P_0$ \\
$W$ & public pretrained weights \\
$L$, $s$ & number of Transformer blocks; split layer \\
$h=\phi_{1:s}(x;W)$ & hidden states computed locally by $P_0$ \\
$\theta=\{A_j,B_j\}_{j=s+1}^{L}$ & LoRA parameters of $P_1$, rank $r$ \\
$\theta_0$, $\theta_T$ & initial and final adapter \\
$b$, $T$, $\eta$ & batch size, number of steps, learning rate \\
$z$, $p$ & logits of the label words; $p=\mathrm{softmax}(z)$ \\
\bottomrule
\end{tabularx}
\end{table}

\end{document}